\documentclass[pre,twocolumn,showpacs,superscriptaddress]{revtex4}

\usepackage{graphicx} 
\usepackage{epsfig}
\usepackage{bm}        
\usepackage{verbatim}
\usepackage{amssymb}   
\usepackage[colorlinks]{hyperref}
\usepackage{verbatim}
\usepackage{mathrsfs}
\usepackage{latexsym} 
\usepackage{bbm}
\usepackage{amsfonts,amsmath}
\usepackage{amssymb}
\usepackage{color}
\usepackage[T1]{fontenc}

\usepackage{eufrak}
\usepackage[colorlinks]{hyperref}
\hypersetup{colorlinks=true,
	linkcolor=red,
	anchorcolor=green,
	citecolor=blue,
	urlcolor=green
}

\providecommand{\be}{\begin{equation}}
\providecommand{\ee}{\end{equation}}
\providecommand{\bea}{\begin{eqnarray}}
\providecommand{\eea}{\end{eqnarray}}
\providecommand{\beas}{\begin{eqnarray*}}
\providecommand{\eeas}{\end{eqnarray*}}

\providecommand{\beni}{\begin{equation*}}
\providecommand{\eeni}{\end{equation*}}

\providecommand{\bw}{\begin{widetext}}
\providecommand{\ew}{\end{widetext}}
 \providecommand{\no}{\nonumber}

\begin{document}

\title{Real space Renormalization Group analysis of a non-mean field spin-glass}

\author{Michele Castellana}
\email{michele.castellana@lptms.u-psud.fr}
\affiliation{LPTMS, CNRS and Universit\'{e} Paris-Sud, UMR8626, B\^{a}t. 100, 91405 Orsay, France} 
\affiliation{Dipartimento di Fisica, Universit\`a di Roma `La Sapienza' , 00185 Rome, Italy}

\pacs{75.10.Nr,64.60.ae,64.70.qj}

\begin{abstract}
A real space Renormalization Group  approach is presented for a non-mean field spin-glass. This approach has been conceived in the effort to develop an alternative method to the Renormalization Group approaches based on the replica method. Indeed, non-perturbative effects in the latter are quite generally out of control, in such a way that these approaches are non-predictive. On the contrary, we show that the real space method developed in this work yields precise predictions for the critical behavior and exponents of the model. 
\end{abstract}
\maketitle

\section{Introduction}

Spin-glasses, structural glasses, and the physical description of their critical properties have interested statistical physicists for several decades.  The mean-field theory of these models   \cite{parisi1979infinite,derrida1980random} provides a physically and mathematically rich picture of their physics and of their critical behavior.  Notwithstanding the great success of such  mean-field theories, real spin-glasses are non-mean field systems, because they have short-range interactions. It follows that these systems cannot be described by mean-field models. Indeed, the generalization of the above mean-field theories to the non-mean field case is an extremely difficult task that has still not been achieved, so that the  development of a predictive and consistent theory of glassy phenomena for real systems is still one of the most  hotly debated  and challenging problems in this domain \cite{FisherHuse86}.

This task is difficult to achieve because the perturbative field-theoretical techniques \cite{wilson1974renormalization,zinnjustin} yielding the Ising model critical exponents with striking agreement with experimental data do not  apply to locally-interacting glassy systems. Indeed, a considerable difficulty in the set-up of a  loop-expansion for a  spin-glass with local interactions, is that the mean-field saddle-point has a very complicated structure \cite{parisi1979infinite,6}, and non-perturbative effects are not under control, in such a way that the properties of the large scale behavior of these systems are still far from being clarified. 

The physical properties of the paramagnetic-ferromagnetic transition emerge clearly in ferromagnetic systems in the original work of Wilson   \cite{wilson1974renormalization}, where one can write a simple Renormalization Group (RG) transformation, describing a flow  under length-scale re-parametrizations. Later on, people  found out that these RG equations are exact  in non-mean field models with power-law ferromagnetic  interactions, built  on hierarchical lattices as  Dyson Hierarchical Model (DHM) \cite{dyson1969existence}. Indeed, in these models one can write an exact RG transformation for the  probability distribution of the magnetization of the system, in such a way that all the relevant physical information on criticality and all the fundamental RG concepts are encoded into this equation, whose solution can be  explicitly built up with the $\epsilon$-expansion technique \cite{collet1978renormalization, collet1977numerical}. Accordingly,  to investigate the RG properties of  non-mean field spin-glasses it is natural to consider spin-glass models built on hierarchical lattices. This study has been done heretofore only for some particular models.  
 On the one hand, models with local interactions on hierarchical lattices built on diamond plaquettes  \cite{berker1},  have been widely studied  in their spin-glass version, and have been shown to  lead  to weakly-frustrated systems even in their mean-field limit  \cite{gardner1984spin}, and so are not a good representative of a  realistic strongly-frustrated spin-glass.
On the other hand, a RG analysis of a different kind of random models on Dyson hierarchical lattice, and of their physical and non-physical infrared (IR) fixed points,  has been done heretofore \cite{theumann1980critical}. Unfortunately, also in these models spins belonging to the same hierarchical block interact each other with the same   \cite{theumann1980critical} random coupling $J$, in such a way that frustration turns out to be relatively weak and they are not a good representative of  a realistic strongly-frustrated system.

The study of such  non-mean field strongly-frustrated spin-glasses is difficult, also because it is hard to identify the correct order parameter and write the resulting RG equations for a function or functional of it without relying on the replica method, which is generally able to make  predictions for the critical exponents only in the mean-field case \cite{6}. 

%

In  this letter we present a real space RG  method for a non-mean field strongly-frustrated spin-glass on hierarchical lattice, the Hierarchical Edwards-Anderson model (HEA) \cite{franz2009overlap}, that does not rely on the replica method. Even if this method does not identify the order parameter of the system,  it  is interesting from the methodological point of view, because it yields a way to implement Kadanoff's \cite{kadanoff1966scaling} block-spin decimation rule in a strongly-frustrated system, and to write the resulting RG equations. In this way, precise predictions on the  critical exponents are obtained. 


 The HEA  is defined as  a system of $2^{k+1}$  spins $S_1,\ldots,S_{2^{k+1}},\, S_i = \pm 1$, with an energy function defined recursively by coupling two systems, say system $1$ and system $2$, of $2^k$ spins
\begin{eqnarray}\label{1} 
H_{k+1}\left[S_1,\ldots,S_{2^{k+1}}\right] = H_{k}^{1}\left[S_1,\ldots,S_{2^{k}}\right]  + \\ \no 
 +H_{k}^{2}\left[S_{2^k+1},\ldots,S_{2^{k+1}}\right]- 2^{-\sigma (k+1)}  \sum_{i=1}^{2^{k}} \sum_{j=2^{k}+1}^{2^{k+1}} J_{ ij} S_i  S_j,
\end{eqnarray}
where $J_{ij}$ are random couplings distributed according to a Gaussian law with zero mean and unit variance, and $H_0[S]=0$. $\sigma$ is a parameter tuning the decay of the interaction strength between spins with distance. It turns out that for $\sigma<1/2$ the thermodynamic limit is ill-defined, because the interaction energy grows with $k$ faster than the volume $2^k$, while for $\sigma>1$ the interaction energy goes to zero for large $k$, and no finite-temperature phase transition occurs. Accordingly, in the following we will take $1/2<\sigma<1$.  In this interval, the model is a non-mean field one, and the mean-field limit is recovered for $\sigma\rightarrow 1/2$ \cite{franz2009overlap}. 

The critical properties of the HEA have been studied heretofore within the replica formalism \cite{castellana2010renormalization}, showing that the system has a classical behavior in the region $1/2<\sigma\leq 2/3$, where the mean-field approximation is correct, while non-mean field effects are important for $2/3<\sigma<1$. This analysis makes a prediction for the critical exponents only in the classical region, since in the non-classical region the first few orders of the $\sigma-2/3 \equiv \epsilon $-expansion have a non-convergent behavior, and higher orders are not known.

Before exposing the real space approach for the HEA, let us illustrate it in the case where the couplings $J_{ij}$ in eq. (\ref{1}) are ferromagnetic, i. e. in for the well-know DHM \cite{dyson1969existence}, in order to test of the consistency of our method.

\section{The real space approach for Dyson Hierarchical Model} \label{real_dyson}

DHM is defined \cite{dyson1969existence} as  a system of $2^{k+1}$  spins $S_1,\ldots,S_{2^{k+1}},\, S_i = \pm 1$, with an energy function defined recursively by coupling two systems, say system $1$ and system $2$, of $2^k$ spins
\begin{eqnarray}\label{20} 
&&  H^ F_{k+1}\left[S_1,\ldots,S_{2^{k+1}}\right] =  H^ F_{k}\left[S_1,\ldots,S_{2^{k}}\right] + \\ \no 
&&+H^ F_{k}\left[S_{2^k+1},\ldots,S_{2^{k+1}}\right] + \\ \no
&& -J 2^{ 2(1-\sigma_F)(k+1)} \Bigg( \frac{1}{2^{k+1}}\sum_{i=1}^{2^{k+1}}  S_i \Bigg)^2  ,
\end{eqnarray}
where $H_0^F[S]=0$,  the suffix $\textrm{F}$ stands for `ferromagnetic', and one can show that $1/2<\sigma_{F} <1$, with the same considerations as those used to derive the constraints on $\sigma$ for the HEA. 

\begin{figure} 
\includegraphics[height=2.5cm,width=5.5cm]{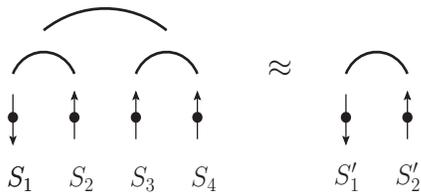}
\caption{Approximation of the real space approach for $k_0=2$. 
In the implementation of the real space approach to DHM, a  $2^2$-spin DHM is approximated by a $2$-spin DHM.
In the implementation of the real space approach to HEA, a  $2^2$-spin HEA is approximated by a $2$-spin HEA.}
\label{fig16}
\end{figure}

The real space RG method is built up by iterating exactly the recursion equation (\ref{20}) for $k=k_0$ steps.
In this way, a DHM model with $2^ {k_0}$ spins $S_1, \ldots, S_{2^{k_0}}$  and Hamiltonian $H_{k_0}^F[S_1, \ldots, S_{2^{k_0}}]$ is obtained. 
We now want to build up a $2^{k_0+1}$-spin DHM starting from such a $2^{k_0}$-spin DHM, which can be done as follows. 
We consider a  $2^{k_0-1}$-spin DHM, where $J$ is replaced by another coupling $J'$. Such a $2^{k_0-1}$-spin DHM is defined by iterating $k_0-1$ times eq. (\ref{20}) with $J \rightarrow J'$, and its Hamiltonian is ${H'}^F_{k_0-1}[S'_1, \cdots, S'_{2^{k_0-1}}]$.  
Given $J$, the coupling $J'$ is chosen in such a way that the $2^{k_0-1}$-spin DHM represents as well as possible the $2^{k_0}$-spin DHM, as qualitatively depicted  in fig. \ref{fig16}. 
 This approximation is practically implemented by considering a physical observable $O^ F_{k_0}(\beta J )$ of the $2^ {k_0}$-spin DHM, and an observable $O^ F_{k_0-1}(\beta  J'  )$ of the $2^ {k_0-1}$-spin DHM, where $\beta$ is the inverse temperature.  The normalized magnetization  on the left  half of the $2^ {k_0}$-spin DHM is 
   \be \no 
  m_{ L} \equiv \Bigg( \frac{1}{2^ {k_0-1}} \sum_{i=1}^ {2^ {k_0-1}} S_i \Bigg)  \Bigg\{\mathbb{E}_{\vec{S}}  \Bigg[ \Bigg( \frac{1}{2^ {k_0-1}} \sum_{i=1}^ {2^ {k_0-1}} S_i    \Bigg) ^2  \Bigg] \Bigg\}^{-\frac{1}{2}},
  \ee
and so for the right-half magnetization $m_R$,   where $\mathbb{E}_{\vec{S}}$ stands for the thermal average at fixed $\beta$, performed with weight $\exp(-\beta H_{k_0}^F)$.  
   The normalized magnetization  on the left-half of the $2^ {k_0-1}$-spin DHM is 
   \be \no 
  m'_{ L} \equiv    \Bigg( \frac{1}{2^ {k_0-2}} \sum_{i=1}^ {2^ {k_0-2}} S'_i \Bigg)       \Bigg\{ \mathbb{E}_{\vec{S'}}  \Bigg[ \Bigg( \frac{1}{2^ {k_0-2}} \sum_{i=1}^ {2^ {k_0-2}} S'_i    \Bigg) ^2  \Bigg]  \Bigg\}^{-\frac{1}{2}}, 
  \ee
 and so for the right-half magnetization $m'_R$, where $\mathbb{E}_{\vec{S'}}$ stands for the thermal average with weight $\exp(-\beta {H'}_{k_0-1}^F)$.
Mimicking Kadanoff's block-spin rule, for the $2^{k_0-1}$-spin DHM to be a good approximation of the $2^{k_0}$-spin DHM, we map the block of the spins in the left-half of the $2^{k_0}$-spin DHM, into the block of the spins in the left-half of the $2^{k_0-1}$-spin DHM, and so for the right-half. 
To do so,  we   choose the observables to be
$ O^ F_{k_0}(\beta J )  \equiv  \mathbb{E}_{\vec{S}} \left[  m_L m_R  \right]$,  
$ O^ F_{k_0-1}(\beta J')  \equiv  \mathbb{E}_{\vec{S}' } \left[  m'_L m'_R \right]$,
and impose the equation
\be \label{rg_equation_ferro}
O^ F_{k_0}( \beta J ) =   O^ F_{k_0-1}(\beta J').
\ee
For any fixed $J$, eq. (\ref{rg_equation_ferro}) determines $J'$ as a function of $J$, as the value of the coupling of the $2^{k_0-1}$-spin DHM such that this yields the best possible approximation of the $2^{k_0}$-spin DHM.

Let us take two copies of the $2^ {k_0-1}$-spin DHM. We put these two copies into interaction according to eq. (\ref{20}), and form a $2^{k_0}$-spin DHM. Since each of the DHMs that we put into interaction represents a $2^{k_0}$-spin DHM, the $2^{k_0}$-spin DHM result of this composition effectively represents a $2^{k_0+1}$-spin DHM. Once again, this DHM is then  approximated as a $2^ {k_0-1}$-spin DHM with coupling, say, $J''$, and so on. Setting $J_0\equiv J, J_1 \equiv J', J_2 \equiv  J'', \ldots$, eq. (\ref{rg_equation_ferro}) establishes a relation between $J_{k}$ and $J_{k+1}$, physically representing the RG flow of the coupling $J_{k}$ under reparametrization of the unit length $2^k \rightarrow 2^{k+1}$.  

The RG eq. (\ref{rg_equation_ferro}) is not exact, because it relies on the fact that a $2^{k_0}$-spin DHM is approximated with a $2^{k_0-1}$-spin DHM. Nevertheless, such an approximation must become asymptotically exact in the large $k_0$-limit, where both $2^{k_0}$ and $2^{k_0-1}$ tend to infinity. 
Quite large values of $k_0$ can be reached by exploiting the hierarchical structure of the system \cite{dyson1969existence}, in such a way that the observables $O^ F_{k_0},O^ F_{k_0-1}$ can be calculated with a computational effort of $2^{k_0}$. It is possible to show that for any $k_0$ the real space method reproduces the constraints $1/2<\sigma_{F} <1$. Indeed, for $\sigma_F > 1$ eq. (\ref{rg_equation_ferro}) gives $J'<J\, \forall J,\beta $, so that the coupling $J_k$ goes to $0$ for large $k$, and no phase transition occurs. 
On the contrary, for $\sigma_F <1/2$ one has $J' > J \, \forall J,\beta$, and the model is thermodynamically unstable. 

\begin{figure}
\includegraphics[width=8.5cm]{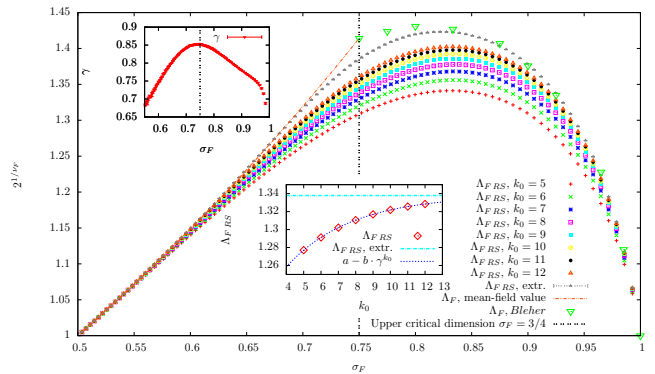}
\caption{
$2^{1/\nu_F}$ as a function of $\sigma_F $ for $1/2\leq \sigma_F\leq 1$. The black dashed line represents the upper   critical dimension  $\sigma_F=3/4$ discussed in \cite{collet1978renormalization}. The points are given by $2^{1/\nu_F}$ computed with the real space method for $5\leq k_0\leq 12$, and the gray points are $2^{1/\nu_F}$ extrapolated to $k_0 \rightarrow \infty$ by fitting $\Lambda_{F \, RS}$ vs $k_0$ with a function of the form $f(k_0) \equiv a- b \cdot \gamma^{k_0}$. The orange dashed curve and the green triangular points are  $2^{1/\nu_F}$ obtained in \cite{bleher1975critical}. Top inset: $\gamma$ vs $\sigma_F$. Bottom inset: $\Lambda_{F \, RS}$ vs $k_0$ for $\sigma_F = 0.92$, its fitting function $f(k_0)$ and the extrapolated value $a$. 
}
\label{fig10}
\end{figure}

The critical exponent $\nu_F$ related to the divergence of the correlation length \cite{zinnjustin} is easily obtained by linearizing the transformation $\beta J\rightarrow \beta J'$ in the neighborhood of the critical fixed-point $\beta J = \beta J' \equiv K_c$ \cite{wilson1974renormalization},
$
2^{1/\nu_F} = \left . \frac{d \beta J'}{d \beta J}  \right|_{\beta J = K_c} . 
$
In fig. \ref{fig10} we depict $2^{1/\nu_F}$ computed with this method, together with  $2^{1/\nu_F}$ computed by Bleher \cite{bleher1975critical,collet1977numerical} with an independent approach, as a function of $1/2\leq \sigma_F\leq 1$.  The latter calculation makes an exact prediction  for $2^{1/\nu_F}$ in the region $1/2<\sigma_F\leq 3/4$ where the mean-field approximation is exact, while it estimates  $2^{1/\nu_F}$ in the non-mean field region $3/4<\sigma_F \leq 1$ by means of a series of successive approximations. 
In the bottom inset of fig. \ref{fig10} we show how $\Lambda_{F \, RS}$ for finite $k_0$ has been extrapolated to the $k_0 \rightarrow \infty$-limit: for every $\sigma_F$ the sequence $\Lambda_{F \, RS}$ vs $k_0$ is fitted with a function of the form $a- b \cdot \gamma^{k_0}$, and $a$ is the resulting extrapolated value. The parameter $\gamma < 1$ is an indicator of the speed of convergence with respect to $k_0$: the larger $\gamma$ the slower the convergence. In the main plot of fig. \ref{fig10} the extrapolated value is depicted, and this is in good agreement with the value given in \cite{bleher1975critical}. The region where the disagreement between the two methods is maximum is $\sigma_F \approx 3/4$, where $2^{1/\nu_F}$ must be  non-analytic \cite{collet1978renormalization}. This non-analyticity cannot show up for finite $k_0$. However, in the top inset of fig. \ref{fig10} we show how the parameter $\gamma$ has a maximum at $\sigma_F \approx 3/4$. This fact shows that the convergence slows down in the proximity of $\sigma_F = 3/4$, i. e. hat the real space method signals the appearance of a non-analyticity of $\nu_F$ at $\sigma_F =3/4$, which results from the switchover from a mean-field to a non-mean field regime.  

It is now  natural to generalize this real space approach to the HEA model, to compare  its predictions with those obtained with the replica method.

\section{The real space approach for the  Hierarchical Edwards-Anderson model} \label{real_hea}

Let us now illustrate how to apply the real space method to the HEA, by considering first the simplest case $k_0=2$.
The reader should follow our derivation in close analogy with that exposed above for DHM. 
 A  HEA model   with $2^ {2}$ spins $S_1,\cdots, S_4$ and Hamiltonian ${H}_2[S_1, \ldots, S_4]$ is built up exactly by means of the recursion equation (\ref{1}). We set $\mathcal{J}_{ij} \equiv 2^{-\sigma}  J_{ij}$, where by definition the couplings  $\{ \mathcal{J}_{ij} \}_{ij}$ are independent identically-distributed random variables, and the probability distribution of each of them will be denoted by $p(\mathcal{J})$. 
Thus, we consider a $2$-spin HEA model, whose Hamiltonian reads ${H'}_1[S'_1, S'_2]  = -   \mathcal{ J}'_{12}   S'_1S'_2$.  
For each realization of the couplings $\{ \mathcal{J}_{ij} \}_{ij}$, we choose $\mathcal{J}'_{12}$ as a function of $\{ \mathcal{J}_{ij} \}_{ij}$  in such a way that the $2$-spin HEA yields the best possible approximation of the $2^2$-spin HEA, as qualitatively depicted in fig. \ref{fig16}.

 In order to do so, let us consider a physical observable $O_{2}(\{\beta  \mathcal{J}_{ij} \}_{ij} )$ of the $2^ 2$-spin HEA, depending on the $6$  couplings $\{ \mathcal{J}_{ij} \}_{ij}$ and $\beta$, and  an observable $O_{1}( \beta \mathcal{J}'_{12}  )$ of the $2$-spin HEA.
Inspired by the fact that the order parameter in the mean-field case is the overlap \cite{parisi1979infinite}, here we build up $O_2$ and $ O_1$ as the thermal average of products of spin overlaps. 
 To build up $O_2$ and $O_1$, consider two \textit{real} replicas $\vec{S}^1, \vec{S}^2$ of the spins of the $2^2$-spin model, and  two real replicas $\vec{S'}^1, \vec{S'}^2$ of the spins of the $2$-spin model. The normalized overlap between $\vec{S}^1$ and $\vec{S}^2$ on the left leaf of the $2^2$-spin HEA  is
   \be \no
Q_L \equiv    \frac{S^ 1_1 S^ 2_1 +S^ 1_2 S^ 2_2}{2}    \Bigg\{ \mathbb{E}_{\vec{S}^1,\vec{S}^2}  \Bigg[ \Bigg( \frac{S^ 1_1 S^2_1 +S^1_2 S^2_2}{2} \Bigg) ^2  \Bigg]  \Bigg\}^{-\frac{1}{2}},
  \ee
and so for the right-leaf overlap $Q_R$,  where $\mathbb{E}_{\vec{S}}$ stands for the thermal average at fixed disorder $\{ \mathcal{J}_{ij} \}_{ij}$ and $\beta$.   The normalized overlap between $\vec{S'}^1$ and $\vec{S'}^2$ on the left leaf of the $2$-spin HEA is $Q'_L  = S'^ 1_1 S'^ 2_1$,  and so for the right-leaf overlap $Q'_R$.  Following Kadanoff's decimation rule, we  map the $2^ 2$-spin HEA into the $2$-spin HEA  by imposing that the spins $S_1, S_2$ correspond to the spin $S'_1$, and that the spins $S_3, S_4$ correspond to the spin $S'_2$. This mapping results in a correspondence between $Q_L$ and $Q'_L$, and between $Q_R$ and $Q'_R$.
By choosing the observables as    
   $ O_{2}(\{ \beta \mathcal{J}_{ij} \} )  \equiv  \mathbb{E}_{\vec{S}^1, \vec{S}^ 2} \left[ Q_L Q_R\right]$, 
   $ O_{1}(\beta \mathcal{J}'_{12})  \equiv  \mathbb{E}_{\vec{S}'^1, \vec{S}'^ 2} \left[  Q'_L Q'_R \right]$, 
Kadanoff's decimation rule  can be practically implemented by imposing the equality   
\be \label{158}
 O_{2}(\{ \beta \mathcal{J}_{ij} \} ) = O_{1}(\beta \mathcal{J}'_{12}),
\ee
where $\mathbb{E}_{\vec{S'}}$ stands for the thermal average  at fixed disorder $\mathcal{J}' _{12}$ and $\beta$.
For any realization of the couplings $\{ \mathcal{J}_{ij} \}_{ij}$, eq. (\ref{158}) determines $\mathcal{J}'_{12}$ as a function of  $\{ \mathcal{J}_{ij} \}_{ij}$ in such a way that the $2$-spin HEA yields the best possible approximation of the $2^2$-spin HEA.  Accordingly, the distribution $p(\mathcal{J})$ induces a distribution of $\mathcal{J}'_{12}$, that we will denote by $p'(\mathcal{J}'_{12})$.
The mapping between $p(\mathcal{J})$ and $p'(\mathcal{J}')$ can be shown to be given by
\bea \label{159}  
p'(\mathcal{J}') &= &\int \Big[ \prod_{i<j}p(\mathcal{J}_{ij})  d \mathcal{J}_{ij} \Big] \frac{1}{2} \times \\ \no 
&& \times\Bigg[ \delta \left(  \mathcal{J}' - \frac{1}{\beta} \operatorname{arctanh}\left(\sqrt{O_{2}(\{ \beta  \mathcal{J}_{ij} \})} \right)    \right) + \\ \no
&& + \delta \left(  \mathcal{J}' + \frac{1}{\beta} \operatorname{arctanh} \left(\sqrt{O_{2}(\{ \beta \mathcal{J}_{ij} \}} )\right)    \right) \Bigg].
\eea  

According to the iterative construction of eq. (\ref{1}), a new HEA is  then constructed by taking two realizations of the $2$-spin HEA. 
Each realization is given by throwing the coupling $\mathcal{J}'$ according to its probability distribution $p'(\mathcal{J}')$. We put these two copies into interaction to form a $2^2$-spin HEA.
Since each of the HEAs that we put into interaction represents a $2^2$-spin HEA, the $2^2$-spin HEA result of this composition effectively represents a $2^3$-spin HEA. At the next step of the iteration, this $2^2$-spin HEA is again  approximated as a $2$-spin HEA with coupling, say, $\mathcal{J}''_{12}$, and the probability distribution $p''(\mathcal{J}''_{12})$ of  $\mathcal{J}''_{12}$ is computed from $p'(\mathcal{J}')$, and so on.  This step is repeated $k$-times, and a system representing  a $2^{2+k}$-spin HEA is obtained.

 Setting $p_0(\mathcal{J})\equiv p(\mathcal{J})$, $p_1(\mathcal{J}) \equiv p'(\mathcal{J})$, $p_2(\mathcal{J}) \equiv  p''(\mathcal{J} ), \ldots$, eq. (\ref{159}) establishes a relation between $p_{k}(\mathcal{J})$ and $p_{k+1}(\mathcal{J})$, physically representing the RG flow of the probability distribution of the  coupling $p_k(\mathcal{J})$ under reparametrization of the unit length $2^k \rightarrow 2^{k+1}$. 


 Eq. (\ref{159}) has been solved by means of the population dynamics algorithm. In population dynamics, one represents the function $p(\mathcal{J})$ as a population of $P \gg 1$ numbers $\{ \mathcal{J}_i \}_{i=1, \ldots, P}$, where each $\mathcal{J}_i$ has been drawn with probability distribution $p(\mathcal{J}_i)$.
The mapping $p(\mathcal{J}) \rightarrow p'(\mathcal{J}')$ given by eq. (\ref{159}) results into a mapping between $ \{ \mathcal{J}_i \}_i$ and the population $\{ \mathcal{J}'_i \}_{i}$ representing $p'(\mathcal{J}')$. 

The structure of the fixed-points of eq. (\ref{159}) has been thus investigated numerically,  showing that there exists a finite value of $\beta = \beta_c$ such that for $\beta < \beta_c$ $p_k(\mathcal{J})$ shrinks to a $\delta(\mathcal{J})$ as $k$ is increased, while  for $\beta > \beta_c$ $p_k(\mathcal{J})$ broadens, i. e. its variance is an ever-increasing function of $k$. 
The physical interpretation of these two temperature regimes is that for $\beta < \beta_c$ $p_k(\mathcal{J})$ flows to the attractive high-temperature fixed-point with $\mathcal{J}=0$ where spins are decorrelated, while for $\beta > \beta_c$ it flows to the attractive low-temperature fixed-point with $\mathcal{J}=\infty$ where spins are strongly-correlated. 
This fact implies that as the temperature is lowered below $T_c=1/\beta_c$ a phase transition occurs, resulting in the appearance of a collective and strongly-interacting behavior of spins in the low-temperature phase. The existence of such a finite-temperature phase transition for a diluted version of HEA model has already been established heretofore in MC simulations by means of finite-size scaling techniques \cite{franz2009overlap}. 

The population dynamics approach  reproduces the fact that for $\sigma<1/2$ the thermodynamic limit is ill-defined, which has been discussed above. Indeed, the numerics show that  for  $\sigma\rightarrow 1/2$ $\beta_c \rightarrow 0$, in such a way that  the variance of $p_k(\mathcal{J})$ , and so that of $H_2$,  is an ever-increasing function of $k$, and the thermodynamic limit $k \rightarrow \infty$ is ill-defined. Unfortunately, the second constraint $\sigma<1$ is not reproduced. This is presumably due to the fact that eq. (\ref{159}) implements only the lowest-order approximation of the real space method, $k_0=2$, and that the method is exact only for large $k_0$. This hypothesis is supported by the estimate of the critical exponents that we will give in the following, suggesting that the closer $\sigma $ to one, the larger the values of $k_0$ needed to have a good estimate of the exact result.   Accordingly, for $\sigma \rightarrow 1$ a significantly better description would be obtained if larger values of $k_0$ were accessible, and the $\sigma<1$-limit would be recovered.

The numerical  implementation  of eq. (\ref{159})  also reveals the existence of a repulsive critical fixed-point with a finite width, that we will denote by $p_\ast(\mathcal J)$, which is reached by iterating  eq. (\ref{159}) with  $\beta = \beta_c$. 
The critical exponent $\nu$ governing the power-law divergence of the correlation length at $\beta = \beta_c$ is determined  \cite{wilson1974renormalization} from the spectrum of the matrix  linearizing the transformation (\ref{159}) in the neighborhood of $p_\ast(\mathcal J)$.

Before discussing the numerical results for $p_\ast(\mathcal J)$ and $\nu$, let us discuss better implementations with $k_0>2$ of this method. The only new element with respect to the $k_0=2$-case is the following. For $k_0>2$, a $2^{k_0}$-spin HEA is approximated as a  $2^{k_0-1}$-spin HEA. The latter has $2^{k_0-1}(2^{k_0-1}-1)/2\equiv M'>1$ couplings $\{ \mathcal{J}'_{ij }\}_{ij}$. It turns out that even if the couplings $\{ \mathcal{J}_{ij }\}_{ij}$ of the $2^{k_0}$-spin HEA are independent, $\{ \mathcal{J}'_{ij }\}_{ij}$ are not, and are distributed according to a joint distribution that we denote  by $p'_{C}(\{ \mathcal{J}'_{ij}\}_{ij})$. In other words, correlations are introduced when iterating once the RG transformation. In the present treatment such correlations have been neglected by assuming that each of the  $\{ \mathcal{J}'_{ij }\}_{ij}$ behaves as an independent random variable distributed according to a distribution obtained as the average of $M'$ distributions, each obtained by marginalizing $p'_{C}(\{ \mathcal{J}'_{ij}\}_{ij})$ with respect to $M'-1$ couplings $\mathcal{J}'_{ij}$. 

\begin{figure}
\includegraphics[width=8.5cm]{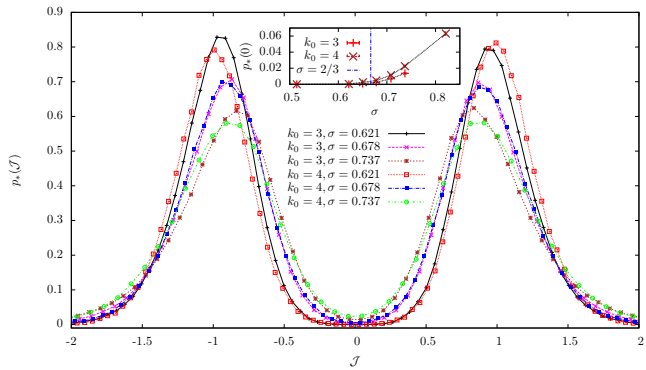}
\caption{Fixed-distribution $p_\ast(\mathcal{J})$ as a function of $\mathcal{J}$ for $k_0=3$ and $\sigma=0.621,0.678,0.737$ (in black, violet and brown respectively), and for $k_0=4$ and $\sigma=0.621,0.678,0.737$ (in red, blue and green respectively). For all the three values of $\sigma$, the discrepancy between  $p_\ast(\mathcal{J})$ in the $k_0=3$-approximation and $p_\ast(\mathcal{J})$ in the $k_0=4$-approximation is relatively small, signaling that $k_0=4$ is presumably large enough for the method to give a reasonably good estimate of the critical fixed-point.
Inset: $p_\ast(0)$ vs $\sigma$ for $k_0=3,4$. A plausible picture resulting from the data is that, for large $k_0$, $p_{\ast}(0) = 0$ for $\sigma < 2/3$ and $p_{\ast}(0) > 0$ for $\sigma > 2/3$. Such picture has a clear physical interpretation given in the text, and suggests a change of behavior at $\sigma=2/3$, reminiscent of the switchover from a mean-field regime for $\sigma<2/3$ to a non-mean field regime for $\sigma>2/3$ predicted by the replica approach. 
 }
\label{fig13}
\end{figure}

The real space approach has been thus implemented for $k_0=2,3,4$. Larger values of $k_0$ were not accessible, since the computational effort scales as $2^{2^{k_0}}$. 
All the qualitative features emerging for $k_0=2$ and discussed above, are preserved for $k_0=3,4$. 
In fig. \ref{fig13} we depict $p_\ast(\mathcal{J})$ as a function of $\mathcal{J}$ for several values of $\sigma$ in the $k_0=3,4$ approximations. Two interesting features emerge from fig. \ref{fig13}. Firstly, the discrepancy between  $p_\ast(\mathcal{J})$ in the $k_0=3$-approximation and $p_\ast(\mathcal{J})$ in the $k_0=4$-approximation is relatively small, signaling that $k_0=4$ is hopefully large enough for the real space approach to give a reasonably good estimate of the critical fixed-point, at least for the values of $\sigma$ considered in fig. \ref{fig13}.
 Secondly, a plausible scenario resulting from the inset of fig. \ref{fig13} is that, for large $k_0$,   $p_\ast(0) = 0$ for $\sigma < 2/3$, while $p_\ast(0) > 0$ for $\sigma > 2/3$. Interestingly, the analysis of the HEA based on the replica approach \cite{franz2009overlap,castellana2010renormalization} predicts a sharp change of behavior from a mean-field regime for $1/2< \sigma\leq2/3$ to a non-mean field regime $2/3<\sigma<1$. 
In the real space approach $\mathcal{J}_{ij}$ is nothing but the effective coupling between spins $S_i$ and $S_j$ of a $2^{k_0}$-spin HEA. At the critical point,  $S_i$ is obtained as the coarse-graining of a group of $2^l, l\gg 1$ spins, which have been progressively decimated and reduced to a single, effective degree of freedom $S_i$, and so for $S_j$. In the mean-field case $\sigma < 2/3$ the model is mean-field, and should thus behave as a fully-connected one.  Accordingly, the $2^l$-spins represented by $S_i$ must interact with all the other spins, and so with the $2^l$-spins representing $S_j$. Thus, the effective coupling between $S_i$ and $S_j$ cannot vanish, i. e. $p_\ast(0)=0$. In the non mean-field case $\sigma > 2/3$ the system is not fully-connected, since the effective interaction range is finite.  Accordingly, there is a finite probability that $2^l$-spins represented by $S_i$ don't interact with the $2^l$-spins representing $S_j$. Thus, the effective coupling between $S_i$ and $S_j$ can vanish, i. e. $p_\ast(0)>0$. 
According to this argument, this change of behavior of $p_\ast(0)$ at $\sigma=2/3$ can be seen as the switchover from a mean-field behavior to a non-mean field one, and is predicted independently and confirmed by the replica analysis of the HEA.

\begin{figure}
\includegraphics[width=8.5cm]{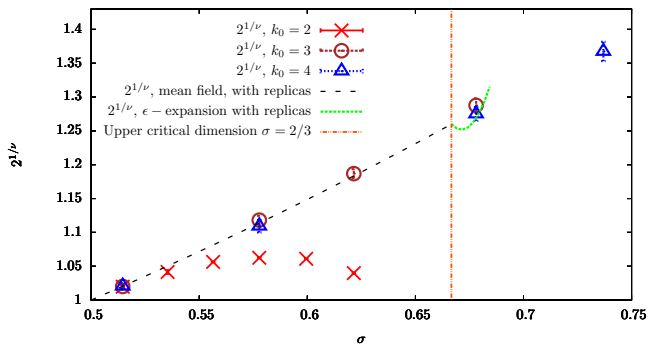}
\caption{
$2^{1/\nu}$ as a function of $\sigma$. The red, brown and blue points are $2^{1/\nu}$  computed with the real space approach for $k_0=2,3,4$ respectively. 
The black  dashed curve and the green dashed curve are $2^{1/\nu}$ obtained with the replica approach \cite{castellana2010renormalization},
and the orange dashed line is the relative upper critical dimension  $\sigma=2/3$ \cite{franz2009overlap}: the black dashed curve is the mean-field value of $2^{1/\nu}$ for $\sigma \leq 2/3$, while the green dashed curve is the two-loops result obtained with the $\epsilon$-expansion. 
}
\label{fig15}
\end{figure}

Let us now consider the predictions on the critical exponent $\nu$. In fig. \ref{fig15} we depict $2^{1/\nu}$ obtained with the  $k_0=2,3,4$-approximation and  $2^{1/\nu}$ from the replica approach  \cite{castellana2010renormalization} as a function of $\sigma$, both in the mean-field region $\sigma\leq 2/3$ and in the non-mean field region $\sigma>2/3$, where the first two orders of the $\epsilon$-expansion are depicted.  
 The agreement between $2^{1/\nu}$ computed with the real space approach for $k_0=2$ and $2^{1/\nu}$ computed with the replica approach is not satisfying. Nevertheless, for  $k_0=3,4$ the agreement in the mean-field region $ 1/2 < \sigma  \leq 2/3$ becomes very good, and serves as an important test of the real space method. A quantitative comparison between $2^{1/\nu}$ of the real space approach and that of the replica approach in the non-mean field region cannot be done, because in the latter the $\epsilon$-expansion is out of control, i. e. the first two orders of the expansion have a non-convergent behavior, and higher orders are not known. Accordingly, the $\epsilon$-expansion curve depicted in fig. \ref{fig15} must not be considered as an estimate of $2^{1/\nu}$. A prediction for $\nu$ in the non-classical region $\sigma>2/3$ for a diluted version \cite{franz2009overlap} of the HEA is given by Monte Carlo (MC) simulations \cite{franz2011unpublished}. According to such a numerical work, $2^{1/\nu}$ is a decreasing function of $\sigma$ in the neighborhood of $\sigma=2/3$, which is in disagreement with the results of the real space approach, fig. \ref{fig15}.  This discrepancy will be discussed in the following.

\section{Discussion and conclusions}\label{conc}

In this letter we developed a real space RG approach for a non-mean field spin-glass, the Hierarchical Edwards-Anderson model (HEA). This approach is innovative with respect to the RG approaches to disordered, strongly frustrated systems developed heretofore that generally rely on the replica method \cite{6}. Indeed, the present approach does not make use of the replica method, which is generally predictive only in the mean-field case, and cannot handle perturbatively fluctuations around the mean-field saddle-point, because these turn out to be out of control \cite{6}. Through a systematic approximation scheme, the present  approach implements Kadanoff's block-spin decimation rule \cite{kadanoff1966scaling} on spins. The implementation of such a decimation rule to a disordered, strongly-frustrated system has not been developed heretofore because of the intrinsic difficulties introduced by frustration, and allows for  an effective reduction of the degrees of freedom of the system.
 Kadanoff's block-spin rule is practically implemented by  approximating  a $2^ {k_0}$-spin HEA as a $2^ {k_0-1}$-spin HEA. Such an approximation is practically performed by imposing that some observables of the $2^ {k_0}$-spin HEA are equal to some corresponding observables of the $2^ {k_0-1}$-spin HEA. For large $k_0$, the method is asymptotically exact, and so its predictions on the critical features of the system. 
The method has been tested in the simplest case of Dyson Hierarchical Model, which is the ferromagnetic version of the HEA, and the  resulting predictions on the critical exponents are in good agreement the results obtained heretofore \cite{bleher1975critical}. 

The method has been then applied to the HEA, and identifies the existence of a phase transition in the system, yielding a prediction on the critical exponent $\nu$ related to the power-law divergence of the correlation length at the critical point. Above the upper critical dimension $\sigma<2/3$, the results for $\nu$ are in very good quantitative agreement with those given by the replica method \cite{castellana2010renormalization} even for small $k_0=3,4$. 
Below the upper-critical dimension $\sigma>2/3$, the $\epsilon$-expansion for the critical exponents performed within the replica method  is not predictive, since the first few orders have a non-convergent behavior, and higher orders are not known. Hence, a quantitative comparison between the real space approach and the replica method is not possible. On the contrary, Monte Carlo (MC) simulations \cite{franz2011unpublished} for a diluted version of the HEA yield a prediction for the critical exponents in this region. These are in disagreement with those of the real space approach. This discrepancy could be due both to the smallness of $k_0$ in the real space approach, or to the non-universality of the exponent $\nu$ when passing from the model on the hierarchical lattice to the diluted model, or to the fact that correlations between the spin couplings have been neglected in the real space approach. Accordingly, the quantitative estimate of $\nu$ below the upper critical dimension is a still untamed issue, which could be suitable for future investigations and developments of the present real space method. 

\acknowledgments
I am glad to thank G. Parisi, M. M\'ezard and S. Franz for extremely useful discussions and suggestions on this work, and A. Decelle for collaborating on the real space method for Dyson Hierarchical Model.  I also acknowledge support from the D. I. computational center of University \textit{Paris Sud}.

\bibliographystyle{plain}
\bibliography{bibliography}

\end{document}